\documentclass[conference]{IEEEtran}
\IEEEoverridecommandlockouts
\usepackage{amsmath,amssymb,amsfonts}
\usepackage{listings}
\usepackage{algorithm}
\usepackage{algorithmicx}
\usepackage[noend]{algpseudocode}
\usepackage{graphicx}
\usepackage{textcomp}
\graphicspath{{pictures/}}
\DeclareGraphicsExtensions{.pdf,.png,.jpg}

\usepackage[inline]{enumitem}
\usepackage{xcolor}
\usepackage{tikz}
\usetikzlibrary{shapes,positioning,calc}
\usepackage[hyphens]{url}
\usepackage[unicode, hidelinks]{hyperref}
\usepackage[hyphenbreaks]{breakurl}
\usepackage[style=ieee, bibencoding=utf8, citestyle=numeric-comp]{biblatex}
\def\BibTeX{{\rm B\kern-.05em{\sc i\kern-.025em b}\kern-.08em
    T\kern-.1667em\lower.7ex\hbox{E}\kern-.125emX}}
\usepackage{booktabs}
\usepackage{colortbl}
\usepackage{multirow}
\usepackage{flushend}
\usepackage{eqparbox}
\newdimen{\algindent}
\algnewcommand\LeftComment[2]{%
\hspace{#1\algindent}$\triangleright$ \eqparbox{COMMENT}{#2} \hfill %
}
\algnewcommand{\algorithmicand}{\textbf{ and }}
\algnewcommand{\algorithmicor}{\textbf{ or }}
\algnewcommand{\OR}{\algorithmicor}
\algnewcommand{\AND}{\algorithmicand}

\begin{document}

\title{Casr-Cluster: Crash Clustering for Linux Applications
\thanks{}}

\author{
\IEEEauthorblockN{
  Georgy Savidov\IEEEauthorrefmark{1}\IEEEauthorrefmark{2} and
  Andrey Fedotov\IEEEauthorrefmark{1}
}
\IEEEauthorblockA{
  \IEEEauthorrefmark{1}Ivannikov Institute for System Programming of the RAS
}
\IEEEauthorblockA{
  \IEEEauthorrefmark{2}Lomonosov Moscow State University
}
Moscow, Russia \\
\{avgor46, fedotoff\}@ispras.ru
}

\maketitle

\begin{tikzpicture}[remember picture, overlay]
\node at ($(current page.south) + (0,0.65in)$) {
\begin{minipage}{\textwidth} \footnotesize
  \copyright~2021 IEEE. Personal use of this material is permitted. Permission
  from IEEE must be obtained for all other uses, in any current or future media,
  including reprinting/republishing this material for advertising or promotional
  purposes, creating new collective works, for resale or redistribution to
  servers or lists, or reuse of any copyrighted component of this work in other
  works.
\end{minipage}
};
\end{tikzpicture}

\begin{abstract}
Crash report analysis is a necessary step before developers begin fixing
errors. Fuzzing or hybrid (with dynamic symbolic execution) fuzzing is often
used in the secure development lifecycle. Modern fuzzers could produce many
crashes and developers do not have enough time to fix them till release date.
There are two approaches that could reduce developers' effort on crash analysis:
crash clustering and crash severity estimation. Crash severity estimation could
help developers to prioritize crashes and close security issues first. Crash
clustering puts similar crash reports in one cluster what could speed up the
analyzing time for all crash reports. In this paper, we focus on crash
clustering. We propose an approach for clustering and deduplicating of crashes
that occurred in  Linux applications. We implement this approach as a tool that
could cluster Casr~\cite{fedotov2020casr} crash reports. We evaluated
our tool on a set of crash reports that was collected from fuzzing results.

\end{abstract}


\section{Introduction}
The number of lines of code in modern software rapidly
increases~\cite{codebase}. Nowadays, security development lifecycle is used in industrial
companies and research institutes~\cite{iso08, howard06, gost16}.
Coverage-guided fuzzing~\cite{serebryany16, sargsyan19, fioraldi20} combined
with dynamic symbolic execution ~\cite{vishnyakov20, poeplau20, poeplau21, borzacchiello21, kuts21}
could produce a large number of crashes that are hard to analyze manually by
developers. Crash clustering and severity estimation~\cite{fedotov2020casr} help developers to spend
less time analyzing and fixing security issues.

There are several approaches to compare crashes. For
manual analysis researchers often look at the address of crash instruction.
One automatic method considers call stack at the time when a program
crashes~\cite{rebucket}. Some methods extract the control flow graph for
comparison~\cite{niskov2018crash, jiang2021igor}.
We focus on developing a clustering algorithm for crash reports that could be
collected in Linux OS. To determine the similarity between crash reports, we used some ideas
outlined in this article~\cite{rebucket}. This method is used for Windows crash
reports. We made several improvements when comparing call stacks with
\texttt{libc abort} call. Also we implemented a crash report deduplication pass,
which could be run before clustering.

The paper is organized as follows. Section~\ref{sec:background} provides an
overview of existing techniques. Section~\ref{sec:design} describes proposed
approach. Section~\ref{sec:implementation} reveals some technical details of
clustering tool. Section~\ref{sec:evaluation} shows the results of the applying clustering
method. Section~\ref{sec:conclusion} concludes the paper.
\section{Background}
\label{sec:background}

There are many approaches for determining the similarity or dissimilarity of
crashes that occur as a result of program execution. We examined the following
ones.

\texttt{1. Apport}~\cite{apport}. A standard ubuntu tool that allows users to
detect crashes and fix them in future versions. This tool also tries to avoid
duplicate error reporting in its databases. To do this, it uses its own
deduplication algorithm. Some signature is generated for the newly arrived crash
report, using call stack to identify duplicates. Then apport creates a
query to the database containing the signatures of the currently known problems.
If the crash is new, it will be added to the apport duplicate database.
Otherwise, the user receives the crash ID and the fixed version number of the
program if its bug has been fixed.

\texttt{2. Igor}~\cite{jiang2021igor}. This crash deduplication tool uses program
execution trace to calculate similarity between two crashes. Igor creates a
Control-Flow Graphs, that describe the execution process of each program. Then
tool applies the Weisfeiler-Lehman Subtree Kernel algorithm to differentiate
tests cases of varying root causes. This algorithm produces a similarity matrix,
that is used in Spectral Clustering method. Igor does not use call stack to
compute the similiraty.

\texttt{3. RETracer}~\cite{retracer}. Microsoft crash triaging tool, that uses
reverse execution combined with static forward analysis to get information about
program failure from partial memory dumps. The tool starts the analysis from
where the corrupted data that caused the crash was last used and analyzes the
instructions in reverse order to the frame where the tainted data was generated
the first time. This frame is blamed guilty of the crash. The program performs
analysis without using execution traces, which significantly complicates the
operation of the algorithm.

\texttt{4.} Method outlined in this article~\cite{niskov2018crash} develops ideas implemented in AFL~\cite{afl}. Tool  builds a control flow graph using AFL-like instrumentation. The metric that is used to determine the similarity of crashes is calculated as the ratio of the number of edges present in one graph and absent in the other to the total number of edges in both graphs. If the graphs are identical, the metric value is 0, and if the graphs do not coincide completely, then 1. Thus, if the crash points coincide and the given metric does not exceed a certain predetermined threshold, then the crashes can be defined as similar.

\texttt{5. ReBucket}~\cite{rebucket}. Crash clustering method proposed by Microsoft
and is based on call stack similarity metric. This method has shown itself well
in practice and we used some similar approaches to calculate the similarity and
clustering.

\section{Design}
\label{sec:design}
The main component of the crash, on the basis of which we determine how much crash reports differ, is
the call stack.

\begin{figure}[!h]
    \center{\includegraphics[scale=0.28]{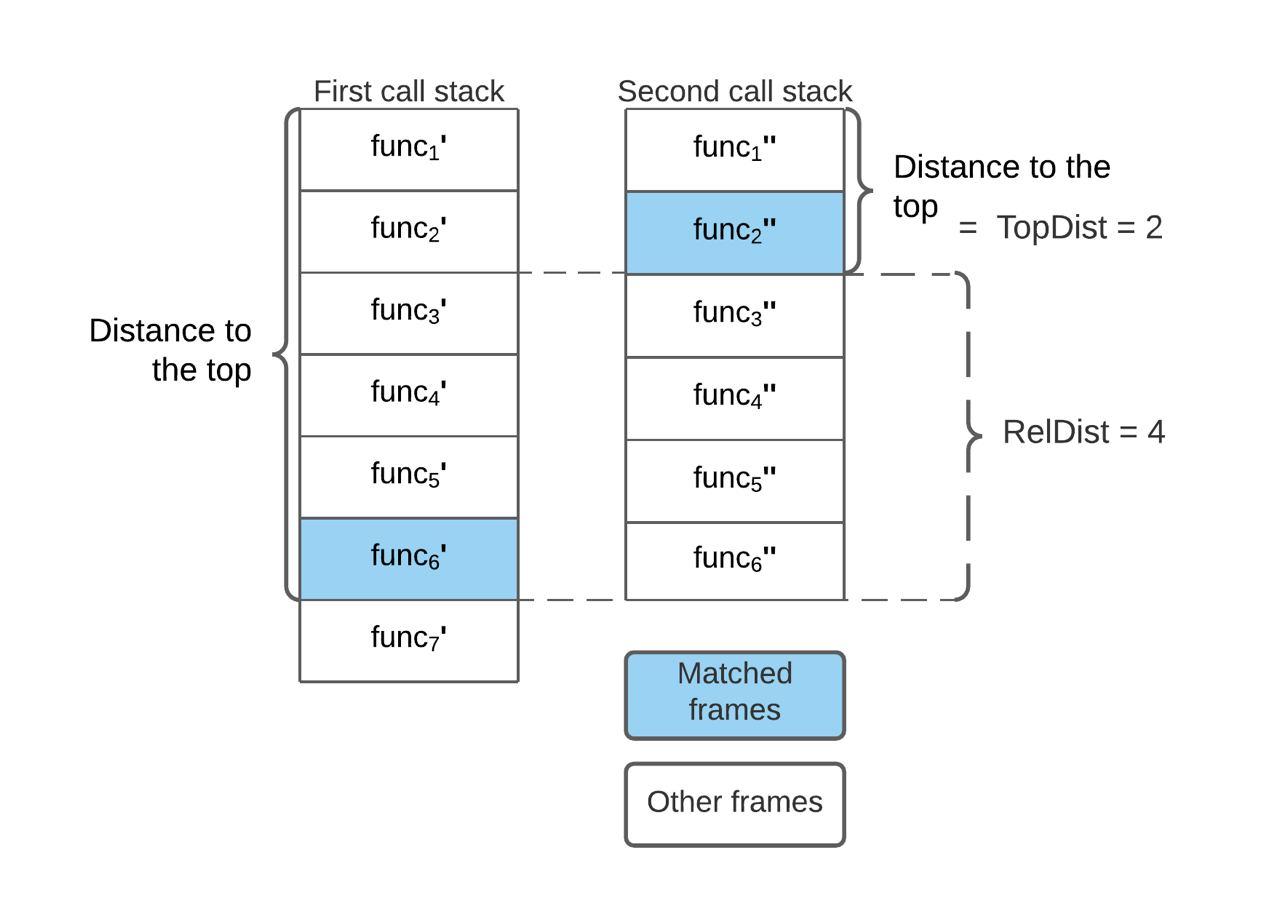}}
    \caption{Call stack metrics illustration.}
    \label{fig:callstack}
\end{figure}

We use the following two metrics in our algorithm(Fig. \ref{fig:callstack}.):

1) Distance from the beginning of the call stacks (\texttt{TopDist}) - the
minimal position offset of the current frame relative to the topmost one.

\texttt{TopDist = min(Distance to the top in first call stack, Distance to the top in second call stack)}

In Fig. \ref{fig:callstack}: \texttt{TopDist = min(6, 2) = 2}.

2) Relative distance (\texttt{RelDist}) - distance between matched frames in two call stacks.

\texttt{RelDist = |Distance to the top in first call stack - Distance to the top in second call stack|}

In Fig. \ref{fig:callstack}: \texttt{RelDist = |6 - 2| = 4}.

Two frames (call sites) are matched if and only if call sites have the same module and the
same offset in this module. We don't use linear address because ASLR is often
present in the system.

Based on these metrics, we make the following assumptions regarding the similarity of the two crashes:

\begin{itemize}
  \item The closer the matching frames are to the top of the call stack, the greater the TopDist weight.
  \item The smaller the distance between the matching frames in call stacks, the greater the weight of RelDist.
\end{itemize}

In our method of calculating the similarity, we also used a dynamic programming
algorithm to find the largest common subsequence of two sequences(Longest Common
Subsequence Problem). Let $f_{1,i}$ the i-th frame in the first call stack and
$f_{2,j}$ the j-th frame in second call stack.We calculate the similarity matrix as follows:
\begin{align}
    \begin{aligned}
        M[i][j] = max
    \begin{cases}
        M[i][j - 1], \\
        M[i - 1][j], \\
        M[i - 1][j - 1] + addition(i,j)
    \end{cases}
\end{aligned}
\end{align}

\begin{align}
    \begin{aligned}
addition(i,j) =
    \begin{cases}
        e^{-r*|i-j|-a*min(i,j)}, &\texttt{$f_{1,i}$ = $f_{2,j}$ } \\
        0 &\texttt{$otherwise$}
    \end{cases} \\
\end{aligned}
\end{align}

$Where:$
\begin{itemize}
    \item \texttt{'r'} - $RelDist$ $coefficient$.
    \item \texttt{'a'} - $TopDist$ $coefficient$.
\end{itemize}

\begin{table*}[h!]
\caption{Casr-cluster Testing Results}
\begin{center}
\scriptsize
    \begin{tabular}{ m{2 cm} m{2 cm} m{2 cm} >{\columncolor{green!15}}m{2 cm} m{2 cm} m{2 cm} m{2 cm} } \toprule
    \textbf{Library} & \textbf{Crash reports} & \textbf{Unique crashes} & \textbf{Number of clusters} &   \textbf{Average number of reports in cluster} &   \textbf{Clustering time(sec)} & \textbf{Deduplication time(sec)} \\
\midrule
	libxml2      & 49     & 10    & 9     & 1     & 5    & 0.08 \\
	jasper       & 231    & 55    & 26    & 2     & 51   & 0.21 \\
	lame         & 74     & 15    & 7     & 2     & 5    & 0.06 \\
	openjpeg     & 264    & 72    & 36    & 2     & 61   & 0.24 \\
	libtiff      & 155    & 56    & 40    & 1     & 21   & 0.14 \\
	libarchive   & 306    & 5     & 3     & 2     & 67   & 0.24 \\
	lrzip        & 38     & 12    & 9     & 1     & 2    & 0.05 \\
	poppler      & 763    & 29    & 15    & 2     & 877  & 1.14 \\
\midrule
    \textbf{TOTAL}: & 1880 & 253 & 154 & 2 & & \\
\bottomrule
\end{tabular}
\label{tbl:casr-cl-res}
\end{center}
\end{table*}

As a result, a certain value will be calculated in the $M[n][m]$(m and n - lengths of 1st and 2nd call stacks) element of the similarity matrix, normalizing which, we will get the similarity of two call stacks (\texttt{similarity\_metric}).

Using \texttt{dist(a, b) = 1 - similarity\_metric (a, b)} as a distance between elements, we apply the hierarchical clustering algorithm, setting previously a certain similarity threshold \texttt{d} as a criterion for stopping the clustering process. Distance between two clusters is defined as

\begin{align}
    \begin{aligned}
        CLdist(CL_i, CL_j) = \underset{a \in CL_i, b \in CL_j}{max(dist(a, b))}
    \end{aligned}
\end{align}
where $CL_i$ and $CL_j$ are different clusters.
As soon as this distance exceeds threshold, the clustering process for this pair of clusters stops. As a result, we get clusters containing similar crash reports.

\begin{figure}[!h]
    \center{\includegraphics[scale=0.3]{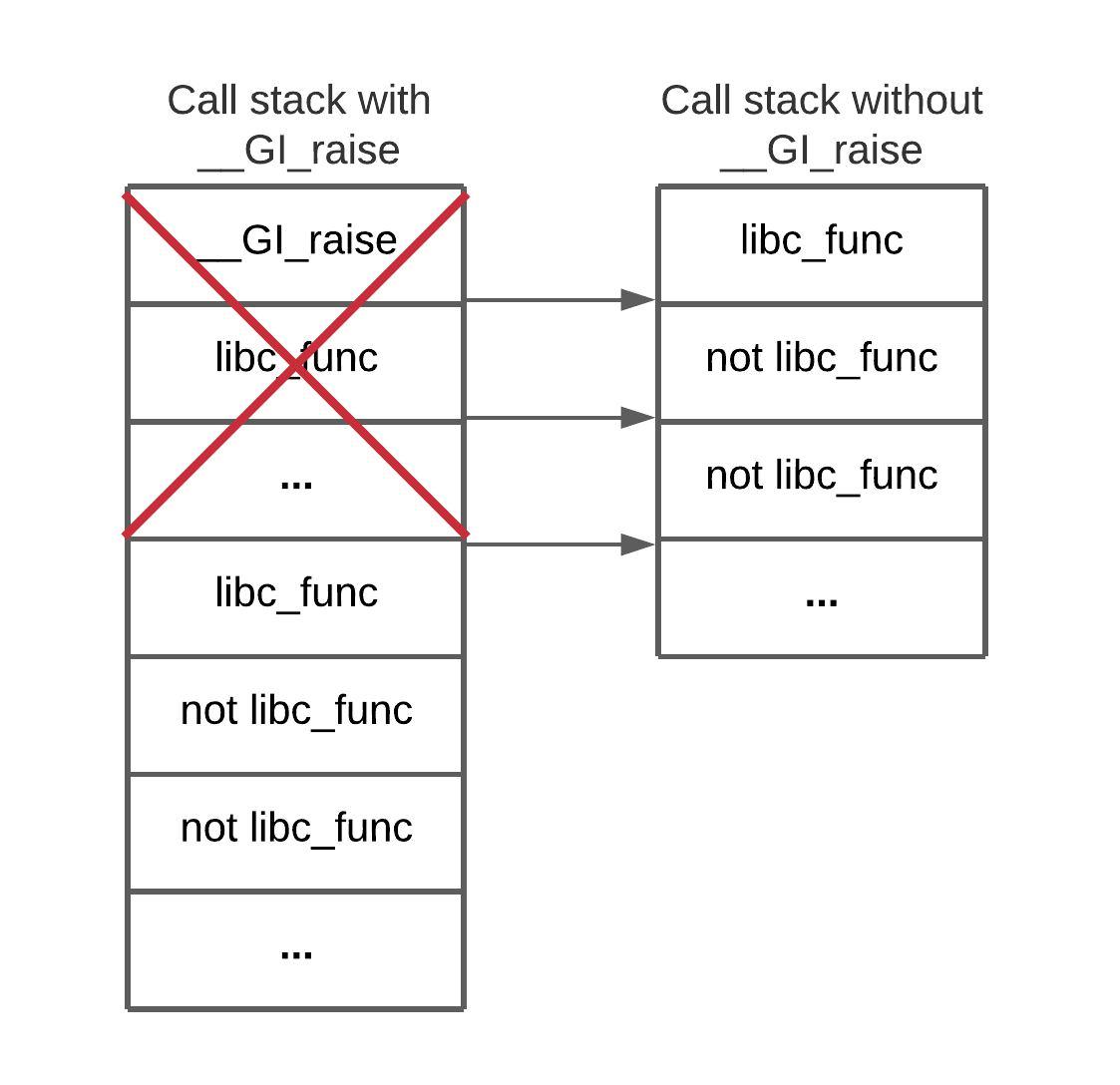}}
    \caption{Removing abort and libc functions from the call stack.}
    \label{fig:abort}
\end{figure}

It is known that with some types of crashes, a long chain of calls to library functions occurs, leading to a call to \texttt{\_\_GI\_raise (abort)}. Often, this chain is the same for different crashes and does not carry any useful information for clustering, but only a noise that should be removed.
We remove the subset of libc library functions from call stack that
lead to the call to \texttt{\_\_GI\_raise}. This subset starts from the top
frame to last libc frame(but not including the last one) before frame that doesn't belong to libc(Fig. \ref{fig:abort}). Such approach gives an essential increase in the accuracy of the algorithm.

We also implemented a  method for reports deduplication. The deduplication
method can be applied to a directory containing reports. Two crashes are considered the same if \texttt{similarity\_metric = 1}. For this, all frames in both Call Stacks must be equal. We can hash each frame (stack trace entry) and the entire Call Stack using attributes such as name of the file to which frames belong and offset from its beginning (information about this is taken from stack trace and mappings). Deduplication utility creates a HashSet and fills it with Call Stacks obtained from reports. As a result, duplicates are removed and only unique reports remain.
\section{Implementation}
\label{sec:implementation}

We have implemented the proposed approach for clustering crash reports based on
Casr~\cite{fedotov2020casr} tool. Casr creates automatic reports for crashes
happened during program testing or deployment. The tool works by analyzing
Linux coredump files. The resulting reports contain the crash’s severity and
additional data (stacktrace, for example) that is helpful for pinpointing the
error cause. So, we implement our clustering tool \texttt{casr-cluster} as a part
of Casr system.

Casr-cluster tool consists of two components. Main module is written in Rust
language and a secondary module in Python. For clustering we use the python library
$scipy.cluster.hierarchy$~\cite{scipy} to launch the hierarchical clustering itself.
In general, the process of the clustering algorithm is as follows:

1) A Rust component calculates pairwise distances between call stacks, writing them
into a condensed distance matrix. To work with call stacks, extracted from casr
reports we use open source library \texttt{gdb-command}~\cite{gdbcommand}. It
allows to compare stacktraces got from gdb or casr reports. Distance matrix is
saved in temporary file.

\begin{figure}[!bh]
    \center{\includegraphics[scale=0.7]{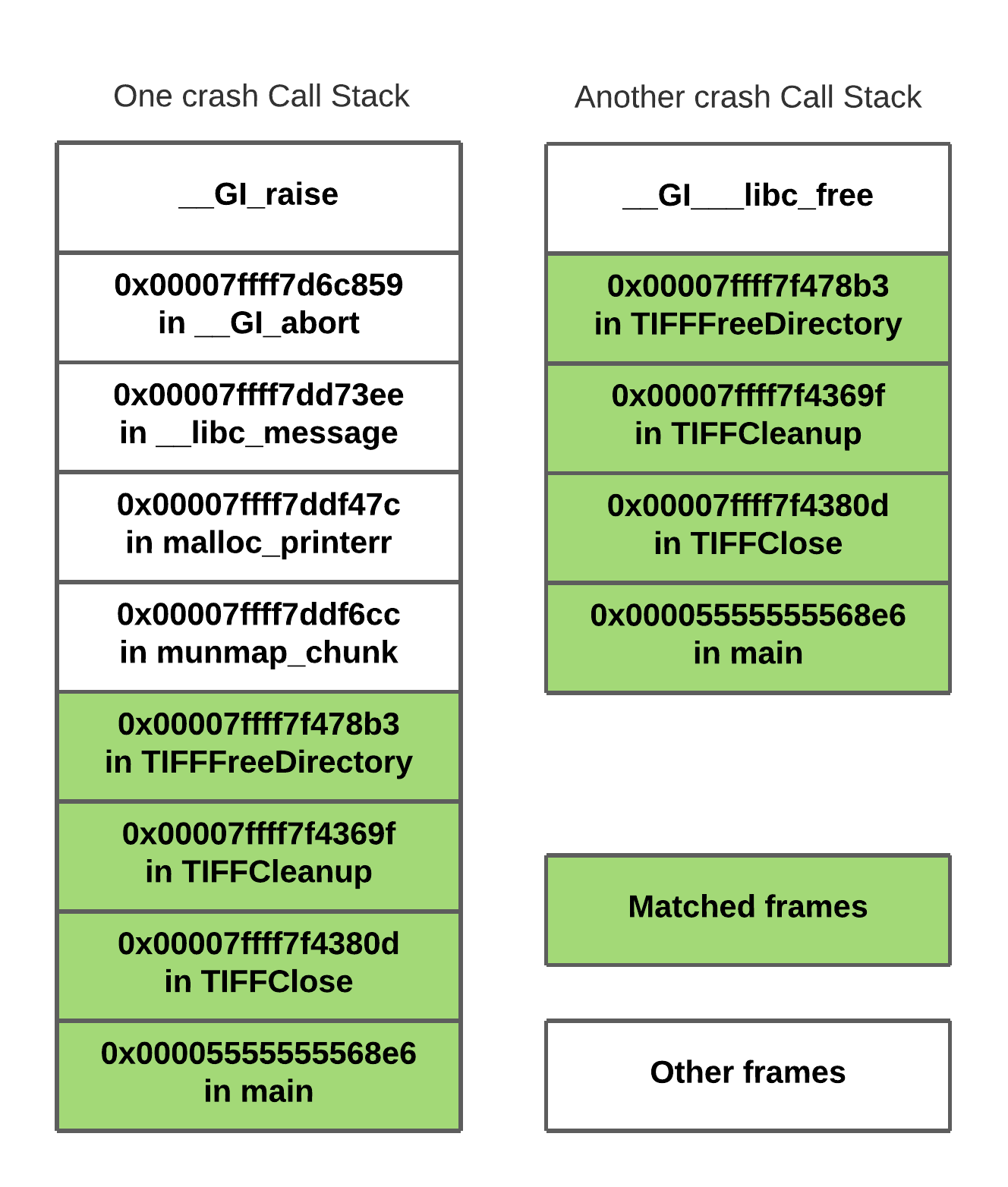}}
    \caption{Two crashes from one cluster.}
    \label{fig:tiff}
\end{figure}

2) The Python module reads a matrix from a temporary file and calls the
hierarchical clustering algorithm ($linkage$) with the $method = 'complete'$
parameter. Then, using the $fcluster$ method, the resulting array
(in which the index is the number of the crash, the value of the element is the
number of the cluster to which the given crash should be assigned) is passed to
the main module.

3) The main module creates the required number of clusters and places crashes
there according to the information received from the array.

\section{Evaluation}
\label{sec:evaluation}

To evaluate the performance and quality of our clustering algorithm, we
collected a dataset consisting of crash reports. We applied fuzzing to some old
versions of the libraries with the known CVEs, which are described in the afl
trophy list~\cite{afltrophy}. As a result, we got a certain set of crashes.
First, we used Casr~\cite{fedotov2020casr} to collect crash reports, then we applied
our clustering algorithm. The results for each of the libraries are described in
the Table \ref{tbl:casr-cl-res} (clustering was performed on a set of
crashes without deduplication). Then we performed deduplication in each cluster.
For better performance report deduplication should be done before clustering,
but some times it is necessary to keep all reports.

Clustering was performed with the coefficients $a = 0.04$, $r = 0.13$ and the threshold value $d = 0.3$.
Thus, it is easy to see that the number of crash reports has decreased by about
an order of magnitude, they were replaced by clusters (with deduplication by
about 1.7 times). Some clusters have more than 3 crashes. In the $poppler$
library, one of the clusters contains 6 similar (but not the same) crashes. Fig.
\ref{fig:tiff}. shows the example of two $libtiff$ crashes that belong to one cluster. We can see that the same error occurred, but different library functions were called to handle it. Thanks to our metrics and \texttt{\_\_GI\_raise} deletion algorithm, crashes were correctly clustered.

The clustering time is at an acceptable level. On a set, containing more than 750 crashes, it did not exceed 15 minutes. Report deduplication time is negligible compared to clustering, even with large sample sizes.

During algorithm testing, we found out that with these coefficients the
program works better when the crashes have a long backtrace. It is necessary to
find the optimal values of the coefficients for the best performance of the
algorithm. This can be done by training the coefficients using machine learning
methods.

\section{Conclusion}
\label{sec:conclusion}

We propose crash clustering method, based on call stack comparison.
Method could be applied to crash reports, collected via Casr tool for Linux
systems. We use optimization for call stack comparison when call stack has libc \texttt{abort}
function call. Also before clustering the report deduplication pass is used. We
evaluated our clustering algorithm on a set of crash reports that was collected
from fuzzing results. The number of crash reports has decreased noticeably.
Crash report clustering could help developers to spend less time analyzing
crashes.

\printbibliography

\end{document}